\def\Granadadep{Departamento de F\'\i sica Te\'orica y del Cosmos,
Facultad
de Ciencias, Universidad de Granada, Campus de Fuentenueva, Granada
18002, Spain}

\def\Granadainst{Instituto de F\'\i sica Te\'orica y Computacional
Carlos I, Facultad
de Ciencias, Universidad de Granada, Campus de Fuentenueva, Granada
18002, Spain}

\def\Valencia{IFIC, Centro Mixto Universidad de Valencia-CSIC,
Burjasot
              46100-Valencia,Spain.}
\def\Comision{Work partially supported by the DGICYT.}

\def\nn{\nonumber}
\def\ni{\noindent}
\def\v{\vskip}
\def\xo{x^0{}}
\def\po{p^0{}}
\def\pomc{\frac{\po}{mc}}
\def\pmc{\frac{p}{mc}}
\def\k{\kappa}
\def\z{\zeta}
\def\Gt{$\widetilde{G}\,\,$}

\def\ant{\vskip 0.4 cm}
\def\desp{\vskip 0.3 cm}

\newcommand{\parcial}[1]{ \frac{\partial}{\partial #1} }

\newcommand{\XL}[1]{ {\tilde{X}}^{L}_{#1} }

\newcommand{\XR}[1]{ {\tilde{X}}^{R}_{#1} }

\newcommand{\iXR}[1]{ i_{\XR{#1}}\Theta}

\documentstyle[12pt]{article}

\begin{document}



\begin{center}
{\Large {\bf FINITE-DIFFERENCE EQUATIONS IN RELATIVISTIC QUANTUM
MECHANICS$^1$ }}
\end{center}

\bigskip
\bigskip

\centerline{ V. Aldaya$^{2,3}$
             and J. Guerrero$^{2,4}$ }     

\bigskip
\centerline{December 14, 1994}
\bigskip

\footnotetext[1]{\Comision}
\footnotetext[2]{\Granadainst} \footnotetext[3]{\Valencia}
\footnotetext[4]{\Granadadep}

\bigskip


\small
\setlength{\baselineskip}{12pt}

\begin{list}{}{\setlength{\leftmargin}{3pc}
\setlength{\rightmargin}{3pc}}
\item Relativistic Quantum Mechanics suffers from structural problems
which are traced
back to the lack of a position operator $\hat{x}$, satisfying
$[\hat{x},\hat{p}]=i\hbar\hat{1}$ with the ordinary momentum operator
$\hat{p}$,
in the basic symmetry group -- the
Poincar\'e group. In this paper we provide a finite-dimensional
extension of the Poincar\'e group containing only one more (in 1+1D)
generator
$\hat{\pi}$, satisfying the commutation relation
$[\hat{k},\hat{\pi}]=i\hbar\hat{1}$
with the ordinary boost generator $\hat{k}$. The unitary irreducible
representations
are calculated and the carrier space proves to be the set of
Shapiro's wave
functions. The generalized equations of motion constitute a simple
example of
exactly solvable finite-difference set of equations associated with
infinite-order polarization equations.
\end{list}

\normalsize
\setlength{\baselineskip}{14pt}


\ant
\ni {\it Introduction: Higher-order Polarizations}
\desp

One essential difference between Geometric Quantization
\cite{GQ1,GQ2,GQ3,GQ4} and a Group
Approach to Quantization (GAQ) \cite{Aldaya82,Loll} is the
possibility of introducing in the
latter higher-order polarizations made out of elements of the (left)
enveloping
algebra. These higher-order polarizations are especially suitable for
those
(anomalous) cases in which the symplectic phase space is not
polarizable, i.e.,
there is no maximal (half the dimension of the manifold) isotropic
distribution of
vector fields with respect to the symplectic form $\omega$, or that
which is the
same, there is no Lagrangian submanifold.

GAQ is formulated on a group \Gt which is a principal bundle with
fibre $U(1)$ and
the symplectic form is replaced by $d\Theta$, where $\Theta$ is the
left
1-form dual to the vertical generator.
The analogous, anomalous problem on a group, corresponds to the
absence
of a first-order full polarization, i.e. a maximal left subalgebra
containing
the kernel of $\Theta$ and excluding the $U(1)$ generator. It can be
solved by
adding operators in the left enveloping algebra to a non-full,
first-order
polarization, thus defining a higher-order polarization \cite{Loll}.
The anomalous
representations of the Virasoro group
related to the non-K\"ahlerian co-adjoint orbits  have
been, for instance, successfully worked out in this way
\cite{Anomalias}.

Higher-order polarizations have also proven to be useful for
representing a physical system in a different (although equivalent)
realization than
the ones given by first-order, full polarizations. This is the case,
for instance,
of the configuration space representation of the free particle and
the
harmonic oscillator (either relativistic or not)
\cite{Position,Oscilata} whose first-order
polarizations lead to momentum space  and Bargmann-Fock
representation,
respectively.

An interesting particularity of the infinite-order polarization is
the
finite-difference character of the generalized equations of motion
and physical
operators, in some sense analogous to the differential realization of
quantum group operators \cite{Mir-Kasimov3,Wolf}. It
constitutes a system  of infinite-order differential operators
closing
a ``weak" algebra: the commutator of two operators only closes, in
general,
on the reduced space of wave functions.

The general form of the solutions to the polarization equations,
leading to
the $q$-realization, can be formally written as:

\begin{equation}
\Psi(t,p,q) \approx e^{-ip\hat{O}_p}e^{-it\hat{O}_t}\psi(q)
\end{equation}

\ni where $t$ represents the variables with non-dynamical character
like time,
rotations, etc., $q$ the generalized coordinates and $p$ their
conjugated momenta. $\psi(q)$  is an arbitrary function
and constitutes the carrier space for the irreducible representation
of the
group. $\hat{O}_t$ is the set of operators generating transformations
in the
variables $t$ (energy, angular momentum, etc.) and $\hat{O}_p$ is a
set of
operators generating (no-local) transformations in $p$, all
of them given in the $q$-realization. When the set
of operators $\{\hat{O}_t,\hat{O}_p\}$ generates (under commutation)
a
subalgebra of the original Lie algebra
the unitarity of the minimal realization in terms of only $q$ and
$\parcial{q}$
is insured, otherwise a further unitarization process is required,
which
consists in a non-canonical choice among algebraicaly equivalent
higher-order
polarizations, i.e. with the
same commutation relations (``symmetrization" process). It must be
stressed that the representation of the group on the complete wave
function
is nevertheless unitary. A general
study of the integration of higher-order polarization will be
published
elsewhere.

In the particular case of Relativistic Quantum Mechanics, the
$p$-realization
(momentum space) is naturally unitary, due to the fact that
$\{\hat{O}_t,\hat{O}_x\}$
generates an algebra isomorphic to the one generated by
$\{\hat{H},\hat{p}\}$,
which is a subalgebra of the Poincar\'e algebra. However, the
$x$-realization
(configuration space) requires further unitarization since
$\{\hat{O}_t,\hat{O}_p\}$ generate an infinite-dimensional algebra
isomorphic
to the one generated by $\{\hat{H},\hat{x}\}$, where $\hat{x}$ is the
position
operator (note that
$[\hat{x},\hat{H}]=i\frac{\hat{p}}{\hat{H}}\approx
i\frac{\hat{p}}{mc} - i\frac{\hat{p}^3}{2m^3c^3} + ...$).

In this letter we elaborate on a finite enlargement of the 1+1-D
Poincar\'e group which
contains a momentum operator $\hat{\pi}$ giving a canonical
commutation relation
with the boost operator $\hat{k}$. A infinite-order polarization on a
central extension
of this group provides a unitary representation of the Poincar\'e
subgroup in a natural way, the support of which are the Shapiro's
wave functions.
It constitutes an example of exactly tractable physical system
associated with
finite-difference (generalized) equations of motion and an
alternative way out
to the problem of the position operator.


\ant
\ni {\it Momentum-space representation of the (1+1D) Poincar\'e
group}
\desp

Let us review very briefly (see \cite{Position} and
references therein for more details)  the standard momentum-space
representation
of the pseudo-extended 1+1D Poincar\'e group on the basis of GAQ. Our
starting point
is the group law for the ordinary (non-extended) Poincar\'e group
$P$.
It is easily derived from its action on the 1+1D Minkowski space-time
parametrized by $\{a^\mu\} \equiv\{a^0,a^1=a\}$:
 $a'^\mu=\Lambda^\mu_{.\nu}(p^0,p)a^\nu + x^\mu$, where $\{x^\mu\}$
are the translations
and $\Lambda$, the boosts, are parametrized by
either p or $\chi\equiv \sinh^{-1}\frac{p}{mc}\equiv
\sinh^{-1}(\gamma
\frac{V}{c}) \equiv 2 \sinh^{-1}\alpha$. $\chi$ is the hyperpolar
co-ordinate parametrizing the (upper sheet of the)
hyperboloid $\po{}^2-p^2=m^2c^2$,
often referred to as the  {\it Lobachevsky space}\
(see \cite{Mir-Kasimov3} and references therein). In terms of $p,\,\,
\Lambda =
\left[\begin{array}{cc} \frac{p^0}{mc}&\frac{p}{mc} \\
 \frac{p}{mc}& 1+\frac{p^2}{mc(p^0+mc)}\end{array}\right]$. As a
manifold,
the group can be seen as the direct product of Minkowski space-time
and
the mass hyperboloid.

The consecutive action of two Poincar\'e transformations leads to the
composition law:
\begin{eqnarray}
\xo'' &=& \xo' + \frac{\po'}{mc}\, \xo + \frac{p'}{mc}\, x \nn \\
x'' &=& x' + \frac{\po'}{mc}\, x + \frac{p'}{mc}\, \xo
\label{Poincare} \\
p'' &=& \frac{\po}{mc}\, p' + \frac{\po'}{mc}\, p \nn
\end{eqnarray}

The Poincar\'e group admits only trivial central extensions by
$U(1)$,
i.e. extensions of the form
\begin{eqnarray}
g''&=&g'*g \qquad g\in P \nn \\
\zeta''&=&\zeta'\zeta e^{i\xi(g',g)} \quad \zeta\in U(1) \nn
\end{eqnarray}

\ni where the {\it cocycle} $\xi$ is a {\it coboundary} generated by
a
function $\eta$ on $P$, $\xi(g',g)=\eta(g'*g)-\eta(g')-\eta(g)$.
We choose $\eta(g)=mcx^0$, so that the $U(1)$ law to be added to
(\ref{Poincare}) is
\begin{equation}
\zeta''=\zeta'\zeta e^{imc(\xo''-\xo'-\xo)} \label{U(1)}
\end{equation}

{}From (\ref{Poincare}) and (\ref{U(1)}) we immediately derive both
left- and right-invariant vector fields:

\begin{eqnarray}
\XL{\xo} &=& \frac{\po}{mc}\parcial{\xo} + \frac{p}{mc}\parcial{x} +
            \frac{\po(\po-mc)}{mc}\, \Xi \nn \\
\XL{x} &=& \pomc\parcial{x} + \pmc\parcial{\xo} +
\frac{p(\po-mc)}{mc}\,\Xi  \\
\XL{p} &=& \pomc\parcial{p} + \pomc\, x\Xi \nn \\
\XL{\z} &=& i\z\parcial{\z} \equiv \Xi \nn
\end{eqnarray}

\begin{eqnarray}
\XR{\xo} &=& \parcial{\xo} \nn \\
\XR{x} &=& \parcial{x} + p\,\Xi  \\
\XR{p} &=& \pomc\parcial{p} + \frac{\xo}{mc}\parcial{x} +
        \frac{x}{mc}\parcial{\xo} + \left( \pmc\,\xo +
\pomc\,x-x\right)\Xi \nn \\
\XR{\z} &=& i\z\parcial{\z} \equiv \Xi \nn
\end{eqnarray}

\ni The pseudo-extended Poincar\'e algebra become

\begin{eqnarray}
\left[\XR{\xo}, \XR{x} \right] &=& 0 \nn \\
\left[\XR{\xo}, \XR{p} \right] &=& \frac{1}{mc}\XR{x}  \\
\left[\XR{x}, \XR{p} \right] &=& \frac{1}{mc}\XR{\xo} - \Xi \nn
\end{eqnarray}

\ni Notice the appearance of the central generator in the third
commutator above, making the extension by $U(1)$ not so trivial  and
justifying the name of pseudo-extension.

The pseudo-extended Poincar\'e group admits a first-order full
polarization which is generated by $<\XL{\xo},\XL{x}>$. The
corresponding
polarized
$U(1)$-functions ($\Xi\Psi=i\Psi$) are
$\Psi = \exp[-i(\po-mc)\xo]\Phi(p)$ and the right generators act on
them
as quantum operators:

\begin{equation}
\begin{array}{rlll}
\hat{p}^0\,\Psi \equiv&i(\XR{\xo}+mc\Xi)\Psi&=\po\Psi
&\Rightarrow\hat{p}^0\,\Phi=\po\,\Phi \\
\hat{p}\,\Psi \equiv&-i\XR{x}\Psi&=p\Psi
&\Rightarrow\hat{p}\,\Phi=p\Phi \\
\hat{k}\,\Psi \equiv&i\XR{p}\Psi&=
i\pomc\,e^{-i(\po-mc)\xo}\frac{\partial\Phi}{\partial p} &\Rightarrow
             \hat{k}\,\Phi= i\pomc \frac{\partial\Phi}{\partial p}
\end{array}
\end{equation}

In this identification of quantum operator with right generators the
rest mass energy has been added to the time generator to obtain the
true energy operator $\hat{p}^0$. This is a consequence  of the fact
that
the pseudo-extension is nothing other than a redefinition of the
$U(1)$
parameter. We must realize that the boosts operator $\hat{k}$ is not
a true position
operator, i.e. it is not $i\frac{\partial}{\partial p}$ and does not
generate
ordinary translations in the spectrum of the momentum operator
$\hat{p}$.

\ant
\ni {\it Relativistic Configuration Space: the S-Poincar\'e group}
\desp

The main problem we face in quantizing Relativistic Mechanics is the
absence
of a commutator like $[\hat{x},\hat{p}]=i\hat{1}$ in the basic
symmetry group,
the Poincar\'e group, where we only find
$[\hat{k},\hat{p}]=i\hat{p}^0/mc$. As
mentioned in the introduction, the
position operator $\hat{x}$ belongs to the infinite-order shell of
the
Poincar\'e algebra and does not close a finite dimensional algebra
with
the rest of the generators. Furthermore, the operator $\hat{p}$ does
not generate
translations on the spectrum of $\hat{k}$ or, equivalently, the
ordinary
Minkowski variable $x$ is not the spectrum of the operator $\hat{k}$.

The solution we propose in this paper is to keep $\hat{k}$ as basic
operator and look for a new momentum operator $\hat{\pi}$  such that
\begin{equation}
[\hat{k},\hat{\pi}]=i\hat{1}\,,\label{kpi}
\end{equation}

\noindent where now $\hat{\pi}$ closes a  finite-dimensional enlarged
Poincar\'e algebra: the S-Poincar\'e algebra. The operator
$\hat{\pi}$
generates true translations on the spectrum of $\hat{k}$, $\k$. The
spectrum of
$\hat{\pi}$, $\pi\equiv mc\chi$, is related to $p$ through
$\pi \equiv mc\sinh^{-1}\pmc = mc\cosh^{-1}\pomc $.
{}.


The group law we propose for the S-Poincar\'e group is given by:
\begin{eqnarray}
\xo'' &=& \xo' + \frac{\po'}{mc}\, \xo + \frac{p'}{mc}\, x \nn \\
x'' &=& x' + \frac{\po'}{mc}\, x + \frac{p'}{mc}\, \xo \nn \\
p'' &=& \frac{\po}{mc}\, p' + \frac{\po'}{mc}\, p \\
\k'' &=& \k' + \k \nn \\
\z'' &=& \z'\z
\exp[imc(\xo''-\xo'-\xo)]\exp[imc\k'\sinh^{-1}\frac{p}{mc}] \nn
\end{eqnarray}

\ni The composition law for the new parameter $\k$ is just additive
and
 does not modify the first three lines (all four lines constituting
 the group law for the non-extended S-Poincar\'e group) since the
associate operator $\hat{\pi}$ commutes with the whole non-extended
Poincar\'e algebra. We can think of $\k$ as parametrizing a
Poincar\'e-invariant space. Furthermore, the extended commutator
(\ref{kpi})
requires a non-trivial cocycle in the composition law for $\zeta$; it
takes the standard form $\exp(i\k'\pi)$. From this group law the
left-invariant vectorfields,
\begin{eqnarray}
\XL{\xo} &=& \frac{\po}{mc}\parcial{\xo} + \frac{p}{mc}\parcial{x} +
            \frac{\po(\po-mc)}{mc}\, \Xi \nn \\
\XL{x} &=& \pomc\parcial{x} + \pmc\parcial{\xo} +
\frac{p(\po-mc)}{mc}\,\Xi \nn \\
\XL{p} &=& \pomc\parcial{p} + ( \k + \pomc\, x)\Xi \\
\XL{\k} &=& \parcial{\k} \nn \\
\XL{\z} &=& i\z\parcial{\z} \equiv \Xi \nn
\end{eqnarray}

\ni and the right ones,
\begin{eqnarray}
\XR{\xo} &=& \parcial{\xo} \nn \\
\XR{x} &=& \parcial{x} + p\,\Xi \nn \\
\XR{p} &=& \pomc\parcial{p} + \frac{\xo}{mc}\parcial{x} +
        \frac{x}{mc}\parcial{\xo} + \left( \pmc\,\xo +
\pomc\,x-x\right)\Xi \\
\XR{\k} &=& \parcial{\k} + mc \sinh^{-1}\pmc\,\,\Xi \nn \\
\XR{\z} &=& i\z\parcial{\z} \equiv \Xi \nn
\end{eqnarray}

\ni are easily derived. The commutators between (say) right
generators are:
\begin{eqnarray}
\left[\XR{\xo}, \XR{x} \right] &=& 0 \nn \\
\left[\XR{\xo}, \XR{p} \right] &=& \frac{1}{mc}\XR{x} \nn \\
\left[\XR{\xo}, \XR{\k} \right] &=& 0 \nn \\
\left[\XR{x}, \XR{p} \right] &=& \frac{1}{mc}\XR{\xo} - \Xi \\
\left[\XR{x}, \XR{\k} \right] &=& 0 \nn \\
\left[\XR{\k}, \XR{p} \right] &=& -\Xi \nn
\end{eqnarray}

The quantization 1-form, i.e. the $U(1)$-left-invariant canonical
1-form,
\begin{eqnarray}
\Theta &=& -(x + \frac{mc}{\po}\,\k) dp - (\po-mc)d\xo +
\frac{d\z}{i\z} \nn\\
       &\equiv& -(x\cosh\frac{\pi}{mc} + \k)d\pi
                -mc(\cosh\frac{\pi}{mc}-1)d\xo + \frac{d\z}{i\z}
\end{eqnarray}

provides the Noether invariants:
\begin{eqnarray}
\iXR{\xo} &=& - (\po-mc) \nn \\
\iXR{x}   &=& p \nn \\
\iXR{p}   &=& -\pomc\,x + \pmc\,\xo -\k \equiv - {\rm K} \\
\iXR{\k}  &=& mc\sinh^{-1}\pmc \equiv \pi \nn
\end{eqnarray}

The integration measure, defined as the product of all left-invariant
forms, is:
\begin{equation}
\Omega = \frac{mc}{\po} d\xo\wedge dx\wedge dp\wedge d\k = d\xo\wedge
dx\wedge d\pi\wedge d\k
\end{equation}

 The characteristic module of $\Theta$ (the Kernel of the Lie algebra
cocycle)
is \linebreak \mbox{${\cal G}_\Theta = < \XL{\xo},\XL{x}-\XL{\k} >$},
and constitutes the generalized
equation of motion. ${\cal G}_\Theta$, or at least a proper
subalgebra, must be
included in any polarization.

\ant
\ni {\it Representations of the S-Poincar\'e group}
\desp

There are three equivalent polarizations which comes out naturally:
\begin{eqnarray}
{\cal P}_p^1 &=& < \XL{\xo},\XL{x}-\XL{\k},\XL{x}+\XL{\k}> \;
             \approx \; < \XL{\xo},\XL{x},\XL{\k}>   \\
{\cal P}_x^{HO} &=& < \XL{\xo} + \left[ \sqrt{m^2c^2-(\XL{x})^2}
-mc\right]\Xi,
                 \XL{\k} +
mc\sinh^{-1}(\frac{i}{mc}\XL{x})\,\Xi,\XL{p} >  \\
{\cal P}_\k^{HO} &=& < \XL{\xo}+mc\left[\cosh (\frac{i}{mc}\XL{\k})
-1\right]\Xi,
                   \XL{x}+mc\sinh (\frac{i}{mc}\XL{\k})\,\Xi,\XL{p} >
\end{eqnarray}

\ni They correspond to the realization of the unitary irreducible
reppresentation in momentum space, $x$-configuration space (ordinary
configuration space) and
$\k$-configuration space, respectively.

\ant
\ni {\it ${\cal P}_p^1$ Polarization: Realization in Momentum Space}
\desp

The polarization equations (once $\z$ has been factorized out
everywhere) give
us the irreducible wave functions:
\begin{equation}
\left. \begin{array}{c}
\XL{\xo}\Psi =0 \\
\XL{x}\Psi =0  \\
\XL{\k}\Psi=0
\end{array} \right\}
\Rightarrow  \Psi = \exp[-i(\po-mc)\xo]\Phi(p)
\end{equation}

\ni on which the right generators act as quantum operators:
\begin{equation}
\begin{array}{rcl}
i\XR{\xo}\Psi &=& (\po-mc)\Psi  \\
-i\XR{x}\Psi &=& p\Psi  \\
i\XR{p}\Psi &=& i\pomc
\exp[-i(\po-mc)\xo]\,\frac{\partial\Phi}{\partial p}  \\
-i\XR{\k}\Psi &=& mc\sinh^{-1}\pmc\,\Psi
\end{array}
\end{equation}

\v 1 cm

\begin{equation}
\Rightarrow\left\{
\begin{array}{rlcl}
\hat{p}^0\,\Phi =& \po\Phi &\equiv& mc\cosh\frac{\pi}{mc}\,\Phi \\
\hat{p}\,\Phi =& p\Phi &\equiv& mc\sinh\frac{\pi}{mc}\,\Phi \\
\hat{k}\,\Phi =& i\pomc\,\frac{\partial\Phi}{\partial p} &\equiv&
              i \frac{\partial\Phi}{\partial \pi} \\
\hat{\pi}\,\Phi =& mc\sinh^{-1}\pmc\,\Phi &\equiv&\pi\,\Phi
\end{array}\right. \label{Spoincaremomento}
\end{equation}

Note that the boosts operator $\hat{k}$ can now be written as
$i \frac{\partial}{\partial \pi}$, i.e. it generates translations in
the
spectrum of $\hat{\pi}$, turning  $\hat{\pi}$ into a ``good" momentum
operator.

\ant
\ni {\it ${\cal P}_x^{HO}$ Polarization: Realization in
$x$-Configuration
Space }
\desp

We follow the same steps as before.
Polarization equations:
\begin{equation}
\XL{p}\Psi =0 \;\Rightarrow \;  \Psi=\exp[-ixp]\exp[-imc\k
\sinh^{-1}\pmc]\,\Phi(\xo,x,\kappa)
\end{equation}

\begin{eqnarray}
& &\left\{\XL{\k} + mc\sinh^{-1}(\frac{i}{mc}\XL{x})\,\Xi\right\}\Psi
=0 \;
\Rightarrow -i\frac{\partial\Phi}{\partial \k} =
  mc\sinh^{-1}(\frac{-i}{mc}\parcial{x})\,\Phi \nn \\
& & \; \Rightarrow
 \Phi(\xo,x,\k) = \exp[imc\k
\sinh^{-1}(\frac{-i}{mc}\parcial{x})]\,\phi(\xo,x)
\end{eqnarray}

\begin{eqnarray}
& &\left\{\XL{\xo} + \left[ \sqrt{m^2c^2-(\XL{x})^2}
-mc\right]\Xi\right\}\,\Psi =0 \;
\Rightarrow i\frac{\partial\phi}{\partial\xo} =
  \left[\sqrt{m^2c^2-\frac{\partial^2}{\partial x^2}}
-mc\right]\,\phi \nn \\
& & \; \Rightarrow
 \phi(\xo,x) = \exp\{-i\xo [\sqrt{m^2c^2-\frac{\partial^2}{\partial
x^2}}-mc]\}\,\varphi(x)
\end{eqnarray}

\begin{eqnarray}
& & \hskip 1 cm \Psi=\exp[-ixp]\exp[-imc\k \sinh^{-1}\pmc]  \nn \\
  & & \times \exp[imc\k \sinh^{-1}(\frac{-i}{mc}\parcial{x})]
  \exp\{-imc\xo [\sqrt{m^2c^2-\frac{\partial^2}{\partial
x^2}}-mc]\}\,\varphi(x)
\end{eqnarray}

Quantum operators (restricted to $\varphi(x)$):
\begin{eqnarray}
\hat{p}^0\varphi &=& \sqrt{m^2c^2-\frac{\partial^2}{\partial
x^2}}\,\varphi  \nn\\
\hat{p}\varphi &=& -i\parcial{x}\,\varphi \nn \\
\hat{k}\varphi &=& \left[
\frac{x}{mc}\sqrt{m^2c^2-\frac{\partial^2}{\partial x^2}} +
                   \k +i\frac{\xo}{mc}\parcial{x}\right]\,\varphi \\
\hat{\pi}\varphi &=&  mc\sinh^{-1}(\frac{-i}{mc}\parcial{x})\,\varphi
\nn
\end{eqnarray}

This realization (restricted to $x$ and $\parcial{x}$) is not
unitary, even though
the representation is unitary on the complete wave functions, because
the $\hat{k}$
operator is not hermitian. This problem is solved by using a proper
(higher-order)
polarization on the Poincar\'e group \cite{Position}. The ${\cal
P}_\k^{HO}$ polarization
on the S-Poincar\'e group provides an alternative solution.

\ant
\ni {\it ${\cal P}_\k^{HO}$ Polarization: Realization in
$\k$-Configuration
Space and the Shapiro Wave Functions}
\desp

Polarization equations:
\begin{equation}
\XL{p}\Psi =0 \;\Rightarrow \;  \Psi=\exp[-ixp]\exp[-imc\k
\sinh^{-1}\pmc]\,\Phi(\xo,x,\kappa)
\end{equation}

\begin{eqnarray}
& &\left\{\XL{\xo}+mc\left[\cosh (\frac{i}{mc}\XL{\k})
-1\right]\Xi\right\}\,\Psi =0 \;
\Rightarrow i\frac{\partial\Phi}{\partial\xo} =
  mc\left[\cosh(\frac{-i}{mc}\parcial{\k}) -1\right]\,\Phi \nn \\
& & \; \Rightarrow
 \Phi(\xo,x,\k) = \exp\{-imc\xo
[\cosh(\frac{-i}{mc}\parcial{\k})-1]\}\,\phi(x,\k)
\end{eqnarray}

\begin{eqnarray}
& &\left\{\XL{x}+mc\sinh (\frac{i}{mc}\XL{\k})\,\Xi\right\}\Psi =0 \;
\Rightarrow -i\frac{\partial\phi}{\partial x} =
  mc\sinh(\frac{-i}{mc}\parcial{\k})\,\phi \nn \\
& & \; \Rightarrow
 \phi(x,\k) = \exp[imcx\sinh(\frac{-i}{mc}\parcial{\k})]\,\varphi(\k)
\end{eqnarray}

\begin{eqnarray}
& & \hskip 1 cm \Psi=\exp[-ixp]\exp[-imc\k \sinh^{-1}\pmc]  \nn \\
  & &  \times\exp\{-imc\xo [\cosh(\frac{-i}{mc}\parcial{\k})-1]\}
\exp[imcx\sinh(\frac{-i}{mc}\parcial{\k})]\,\varphi(\k)
\end{eqnarray}

Quantum operators:
\begin{eqnarray}
\hat{p}^0\varphi &=& mc \cosh(\frac{-i}{mc}\parcial{\k})\,\varphi
\nn\\
\hat{p}\varphi &=& mc \sinh(\frac{-i}{mc}\parcial{\k})\,\varphi \nn
\\
\hat{k}\varphi &=& \k\,\varphi -\xo
\sinh(\frac{-i}{mc}\parcial{\k})\,\varphi
                    + x\cosh(\frac{-i}{mc}\parcial{\k})\,\varphi  \\
\hat{\pi}\varphi &=& -i \parcial{\k}\,\varphi \nn
\end{eqnarray}

As can be seen, this representation of the S-Poincar\'e group, or
more
precisely the one which appears after dropping the $\xo$ and $x$
``evolution"
from the $\hat{k}$ operator, is unitary and
contains a unitary and irreducible representation of the Poincar\'e
subgroup.

An analogous commet to that made under (\ref{Spoincaremomento}) also
apply here, or, the other way round the new parameter $\k$ allows
$\hat{k}$
to be written as a multiplicative operator.

The Shapiro wave functions are nothing other than the eigen-functions
of the momentum operator $\hat{p}$ in
$\k$-configuration space:
\begin{eqnarray}
& &\hat{p}\varphi_p(\k) = p\varphi_p(\k)
\Rightarrow mc\sinh(\frac{-i}{mc}\parcial{\k})\,\varphi_p=p\varphi_p
\nn \\
&\Rightarrow& \varphi_p(\k) = \exp[imc\k \sinh^{-1}\pmc] \equiv
\exp[i\k\pi] =
\left(\frac{\po-p}{mc}\right)^{-imc\k}
\end{eqnarray}


\begin{thebibliography}{99}

 \bibitem{GQ1}           J.M. Souriau, {\it Structure des systemes
                         dynamiques},
                         Dunod, Paris (1970)
 \bibitem{GQ2}           B. Kostant, {\it Quantization and Unitary
                         Representations},
                         in Lecture Notes in Math. {\bf 170},
                         Springer-Verlag, Berlin (1970)
 \bibitem{GQ3}           J. Sniatycki, {\it Geometric Quantization
                         and Quantum Mechanics},
                         Springer-Verlag, New York (1970)
 \bibitem{GQ4}           N. Woodhouse, {\it Geometric Quantization},
                         Clarendon, Oxford (1980).
 \bibitem{Aldaya82}      V. Aldaya and J.A.de Azc\'arraga, J. Math.
                         Phys. {\bf 23}, 1297 (1982)
 \bibitem{Loll}          V. Aldaya, J. Bisquert, R. Loll and J.
                         Navarro-Salas, J. Math. Phys. {\bf 33}, 3087
(1992)


\bibitem{Anomalias}     V. Aldaya and J. Navarro-Salas, Commun.
                         Math. Phys. {\bf 139}, 433 (1991)

\bibitem{Position} V. Aldaya, J. Bisquert, J. Guerrero and J.
Navarro-Salas,
                     J. Phys. {\bf A 26}, 5375  (1993)

\bibitem{Oscilata} V. Aldaya, J. Bisquert, J. Guerrero and J.
Navarro-Salas,
                       {\it Group-Theoretical Construction of the
Quantum
                        Relativistic Harmonic Oscillator},
                        Preprint UG-FT-34/93 (1993)



\bibitem{Mir-Kasimov3} R.M. Mir-Kasimov, J. Phys. {\bf A 24}, 4283
(1991)

\bibitem{Wolf} N. Atakishiyev, A. Frank, and K. B. Wolf,
               J. of Math. Phys. {\bf 35}, 3253 (1994)




\end{thebibliography}
\end{document}